\documentclass[sigconf]{acmart}
\acmConference[ESEC/FSE 2023]{The 31st ACM Joint European Software Engineering Conference and Symposium on the Foundations of Software Engineering}{11 - 17 November, 2023}{San Francisco, USA}

\setcopyright{acmcopyright}
\copyrightyear{2023}
\acmYear{2023}
\acmDOI{XXXXXXX.XXXXXXX}

\acmPrice{15.00}
\acmISBN{978-1-4503-XXXX-X/18/06}

\usepackage{amsmath}
\usepackage[normalem]{ulem}
\usepackage{graphicx}
\usepackage{pifont} 
\usepackage{booktabs} 
\usepackage{tabularx}
\usepackage{listings}
\usepackage{longtable}
\usepackage{multicol}
\usepackage{multirow}
\usepackage{hyperref}
\hypersetup{draft} 
\usepackage{caption}
\usepackage{subfig}
\usepackage{stfloats} 
\usepackage{adjustbox}
\usepackage{float}
\usepackage{pgfplotstable}
\usepackage{tikz}
\usepackage{xcolor}
\colorlet{mcadError}{lightgray!60!white}
\colorlet{mcadGMeanError}{red}
\colorlet{mcadDTPMcad}{orange}
\colorlet{mcadDTPPerf}{blue}
\usepackage{titlesec}

\newcommand{\cmark}{\ding{51}}%
\newcommand{\xmark}{\ding{55}}%

\newcommand{\coolname}{MCAD}

\newcommand{\totalError}{3\%}
\newcommand{\ffmpegIntelError}{1.8\%}
\newcommand{\ffmpegAmdError}{2.6\%}
\newcommand{\ffmpegArmError}{1.1\%}
\newcommand{\ffmpegErrorStd}{6.3\%}
\newcommand{\clangIntelError}{2.3\%}
\newcommand{\clangAmdError}{2.2\%}
\newcommand{\clangArmError}{1.9\%}
\newcommand{\clangErrorStd}{5\%}

\begin{document}

\titlespacing*\section{0pt}{3pt plus 0pt minus 1pt}{3pt plus 0pt minus 1pt}
\titlespacing*\subsection{0pt}{2pt plus 1pt minus 1pt}{2pt plus 1pt minus 1pt}
\titlespacing*\subsubsection{0pt}{2pt plus 0pt minus 1pt}{2pt plus 0pt minus 1pt}

\setlength{\abovedisplayskip}{2pt}
\setlength{\belowdisplayskip}{2pt}

\setlength{\floatsep}{3.0pt plus 0pt minus 3.0pt}
\setlength{\textfloatsep}{3.0pt plus 0pt minus 3.0pt}
\setlength{\intextsep}{3.0pt plus 0pt minus 3.0pt}

\title[A Highly Scalable, Hybrid, Cross-Platform Differential Throughput Estimation Framework]{
  A Highly Scalable, Hybrid, Cross-Platform Timing Analysis Framework Providing Accurate Differential Throughput Estimation via Instruction-Level Tracing}

\author{Min-Yih Hsu}
\email{minyihh@uci.edu}
\affiliation{%
  \institution{University of California, Irvine}
  \state{California}
  \country{USA}
}

\author{Felicitas Hetzelt}
\email{fhetzelt@uci.edu}
\authornote{Both authors contributed equally to this research.}
\affiliation{%
  \institution{University of California, Irvine}
  \state{California}
  \country{USA}
}

\author{David Gens}
\email{dgens@uci.edu}
\authornotemark[1]
\authornote{Now affiliated with Cerebras}
\affiliation{%
  \institution{University of California, Irvine}
  \state{California}
  \country{USA}
}

\author{Michael Maitland}
\email{michael.maitland@sifive.com}
\affiliation{%
  \institution{SiFive}
  \state{California}
  \country{USA}
}

\author{Michael Franz}
\email{franz@uci.edu}
\affiliation{%
  \institution{University of California, Irvine}
  \state{California}
  \country{USA}
}

\begin{abstract}

Differential throughput estimation, i.e., predicting the performance impact of software changes, is critical when developing applications that rely on 
accurate timing bounds, such as automotive, avionic, or industrial control systems.
However, developers often lack access to the target hardware to perform on-device measurements, and hence rely on
instruction throughput estimation tools to evaluate performance impacts.

State-of-the-art techniques broadly fall into two categories: dynamic and static.
Dynamic approaches emulate program execution using cycle-accurate microarchitectural simulators
resulting in high precision at the cost of long turnaround times and convoluted setups. 
Static approaches reduce overhead by predicting cycle counts outside of a concrete runtime environment.
However, they are limited by the lack of dynamic runtime information and mostly focus on predictions over single basic blocks which requires developers to manually construct critical instruction sequences.

We present \coolname{}\footnote{The actual name has been redacted for anonymous review.}, a hybrid timing analysis framework that combines the advantages of dynamic and static approaches.
Instead of relying on heavyweight cycle-accurate emulation,
\coolname{} collects instruction traces along with dynamic runtime information from QEMU and streams them to a static throughput estimator.
This allows developers to accurately estimate the performance impact of software changes for complete programs within minutes, reducing turnaround times by orders of magnitude compared to existing approaches with similar accuracy.
Our evaluation shows that \coolname{} scales to real-world applications such as FFmpeg and Clang with millions of instructions, achieving $<$~\totalError{} geo.~mean error compared to ground truth timings from hardware-performance counters on x86 and~ARM machines.

\end{abstract}





\maketitle

\section{Introduction}
Semantically equivalent modifications of a given piece of software can result in varying degrees of performance degradation due to resource contentions on the architectural and microarchitectural level.
For systems that have tight timing restrictions, it is therefore critical to identify specific implementations that minimize negative performance impacts and maintain timing restrictions over the execution of the whole program.
If the target system is not available to perform on-device measurements, developers instead need to rely on tools to estimate cycle counts for a given program or instruction sequence.
To that end, several approaches have been developed that
roughly fall into one of two categories: (i) emulating the execution of concrete runtime instances of the program on simulated hardware (i.e., \emph{dynamic approaches}) and (ii) estimating the cycle count of program instructions without concrete execution under an abstract runtime environment (i.e., \emph{static approaches}).


Dynamic approaches achieve high precision 
using architectural simulators~\cite{burger1997simplescalar, yourst2007ptlsim, binkert2011gem5} which faithfully model the runtime behavior. 
They provide concrete and generally accurate estimates, however, they suffer from high runtime overhead and can require hours to days to analyze a program.
Furthermore, architectural simulators exhibit a high architecture dependence and are complicated to set up,
often requiring dedicated expert knowledge and/or giving rise to compatibility issues with standard tools and default environments.

Static approaches~\cite{laukemann2018automated,mendis2019ithemal,uica,llvmmca, sehlberg2006static, li2007chronos, falk2009optimal, lisper2014sweet, hardy2017heptane} alleviate the performance and setup cost of dynamic hardware simulators.
Traditionally, such approaches target worst-case execution time predictions over all possible execution paths~\cite{sehlberg2006static, li2007chronos, falk2009optimal, lisper2014sweet, hardy2017heptane}.
More recent works~\cite{laukemann2018automated,mendis2019ithemal,uica,llvmmca} focus on smaller execution sequences and also construct parametric models that generalize to multiple architectures.
These models are either trained end-to-end using throughput data~\cite{mendis2019ithemal} or programmatically tuned for key parameters that are publicly available or obtained from measurements~\cite{laukemann2018automated, uica}.
However, static approaches are fundamentally limited in practice due to their lack of concrete dynamic runtime information.
Hence, traditional static approaches lack support for essential program constructs such as loops, data-dependent control flows, and memory accesses~\cite{mezzetti2011industrial, abella2015wcet}.
In addition, the scope of many static throughput estimation tools is limited to throughput predictions of individual basic blocks~\cite{laukemann2018automated,mendis2019ithemal,uica}, i.e., only a handful of instructions.
Even for tools that can in theory process multiple basic blocks at once, predictions do not usually hold across control-flow transfers.
For example, MCA~\cite{llvmmca} does not follow call or jump targets and instead simply falls through to the next instruction while adding a static cycle penalty\footnote{\url{https://github.com/llvm/llvm-project/blob/main/llvm/lib/MCA/InstrBuilder.cpp\#L224}}.

In this paper, we present \coolname{}, a lightweight alternative that provides whole-program throughput prediction of binary software.
\coolname{} follows a hybrid approach that supplements static throughput estimates with dynamic runtime information.
To avoid the overhead of cycle accurate architectural simulation, \coolname{} uses an emulation-based approach to obtain execution traces using  QEMU~\cite{bellard2005qemu} and forwards them to the LLVM Machine Code Analyzer~(MCA)~\cite{llvmmca} for instruction-level analysis.
In addition, \coolname{} extends MCA to resolve several inherent limitations of static throughput estimation.
First MCA's instruction analysis is redesigned to process instructions in a streaming fashion, which enables the analysis to scale to large real-world binaries.
Second, \coolname{} provides an interface to incorporate dynamic information into the analysis.
The dynamic information captures concrete control-flow which allows \coolname{} to accurately predict instruction cycle counts across basic block boundaries.
In addition, the instruction stream can be supplemented with arbitrary metadata, such as memory aliasing properties or execution context information, to further improve prediction accuracy.

The main purpose of \coolname{} is to provide fast, yet accurate differential timing analyses: cycle counts for whole program execution traces which typically contain hundreds of thousands to millions of instructions can usually be produced within the order of minutes or seconds,
which allows developers to quickly identify the least intrusive change with respect to execution time.
We extensively test and evaluate \coolname{} with respect to scalability and accuracy on a number of different real-world applications such as FFmpeg and Clang to demonstrate that \coolname{} can model microarchitectural behaviors, such as instruction latencies in superscalar processors, accurately and with low cost.
The geo.~mean error in differential timing between \coolname{} and hardware performance counters in our experiments is smaller than \totalError{} across several different microarchitectures and application software.

\paragraph{Summary of Contributions:}
\begin{itemize}
\item We present \coolname{}: a new open-source\footnote{Will be made available after de-anonymization. Some of our contributions in this work have been adopted and are currently in use as part of LLVM.} framework for throughput estimation yielding highly accurate differential timings, on par or better than the current state-of-the-art, while reducing turnaround time by several orders of magnitude.
\item Our prototype implementation leverages QEMU as a fast instruction executor, utilizing MCA to model individual per-instruction execution cycles, rather than simulation-based approaches that faithfully model complex processor front-ends.
\item We provide a detailed evaluation of \coolname{} with respect to accuracy and scalability for the popular x86 and ARMv8 instruction-set architectures using several different devices and hardware-performance counters to collect timing measurements for real-world traces as ground truth.
\end{itemize}

\section{Background and Motivation}\label{sec:background_motivation}
In this section we provide background on inherent limitations of static throughput estimation approaches and present a use case scenario to motivate the design goals of \coolname{}.

\subsection{Static Throughput Estimation Challenges}\label{sec:background}

Throughput estimation is an active area of research that aims to statically predict the performance upper bound of a program, usually measured by cycle counts or Instruction Per Cycle~(IPC), of a \textit{single} basic block.
Current tools model microarchitectural details such as instruction latency and number of micro-ops of the target processor.
However, there are two major issues with this approach: (i) it does not easily transfer across branch instructions or function call boundaries (ii) dynamic information such as execution context and memory aliasing is usually not taken into account.

\subsubsection*{Control Flow Transfers.~}
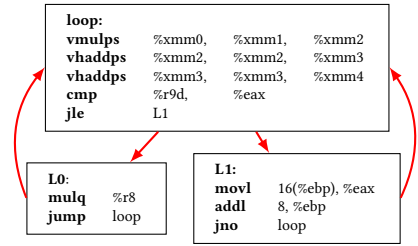
\begin{figure}
    \centering
    \usetikzlibrary{automata} 
\usetikzlibrary{positioning} 
\usetikzlibrary{arrows} 
\tikzset{node distance=1.5cm, 
every state/.style={ 
semithick,
fill=gray!10},
initial text={}, 
double distance=2pt, 
every edge/.style={ 
    draw,
    red,
->, 
auto,
thick}}
\let\epsilon\varepsilon
\tikzset{>=latex}

\begin{tikzpicture}
\node[draw, font=\scriptsize, align=left, anchor=south] (L) {
\begin{tabular}{llll}
    \textbf{loop:} \\
    \textbf{vmulps}  & \%xmm0, & \%xmm1, & \%xmm2\\
    \textbf{vhaddps} & \%xmm2, & \%xmm2, & \%xmm3\\
    \textbf{vhaddps} & \%xmm3, & \%xmm3, & \%xmm4\\
    \textbf{cmp}     & \%r9d,  & \%eax\\
    \textbf{jle}     & L1\\
\end{tabular}
};

\node[draw, font=\scriptsize, align=left, anchor=north west, below left of=L, yshift=-5ex, xshift=-4ex] (L0) {
\begin{tabular}{llll}
    \textbf{L0}: \\
    \textbf{mulq}     & \%r8 \\
    \textbf{jump}     & loop \\
\end{tabular}
};

\node[draw, font=\scriptsize, align=left, anchor=north west, below right of=L, yshift=-5ex] (L1) {
\begin{tabular}{llll}
    \textbf{L1:} \\
    \textbf{movl}     & 16(\%ebp), \%eax \\
    \textbf{addl}     & 8, \%ebp \\
    \textbf{jno}      & loop \\
\end{tabular}
};

    \draw (L) edge node {} (L0);
    \draw (L) edge node {} (L1);
    \draw (L0.mid west) edge[bend left] node {}  (L.mid west);
    \draw (L1.mid east) edge[bend right] node {} (L.mid east);
\end{tikzpicture}
    \caption{x86\_64 assembly control flow.}
    \label{fig:motivate-example}
\end{figure}

Figure~\ref{fig:motivate-example} shows the control flow of an x86\_64 assembly code snippet consisting of three basic blocks, \texttt{loop}, \texttt{L0}, and \texttt{L1}.
Block \texttt{loop} calculates a vector dot product followed by a conditional branch into either block \texttt{L0} or \texttt{L1} based on a data-dependent comparison, with both blocks jumping back to \texttt{loop} at the end.

\lstdefinelanguage{x86asm}{
    keywords = {vmulps, vhaddps, cmp, jle, mulq, addl, addq, movl, movq, jmp, jno}
}
%
%
Using Intel Coffee Lake as an example target architecture, throughput predictions of the individual basic blocks of this program will not generalize across executions.
The reason is that instruction-level throughput for this program actually depends on the ordering of executed basic blocks on that architecture.
In particular, executions where \texttt{L0} follows \texttt{loop}~(Listing~\ref{lst:motivate-example-ordering1}) are roughly 5$\sim$10 cycles slower, per iteration, than executions in which \texttt{L1} follows \texttt{loop}~(Listing~\ref{lst:motivate-example-ordering2}).
This might seem counterintuitive as Listing~\ref{lst:motivate-example-ordering1} contains fewer instructions than Listing~\ref{lst:motivate-example-ordering2} and, more importantly, there is a memory read instruction (\texttt{movl     16(\%ebp), \%eax}) in the latter trace, which should be slower than scalar multiplication (\texttt{mulq     \%r8}).
However, measurements on the target architecture reveal that there is a substantial slowdown in traces that follow the shorter Listing~\ref{lst:motivate-example-ordering1} due to resource contention between the two basic blocks.

\begin{lstlisting}[language=x86asm, caption={Trace of executing \texttt{L0} after \texttt{loop} in Figure~\ref{fig:motivate-example}}, label={lst:motivate-example-ordering1}, captionpos=b, numbers=left, xleftmargin=5.0ex]
    vmulps   %xmm0, %xmm1, %xmm2
    vhaddps  %xmm2, %xmm2, %xmm3
    vhaddps  %xmm3, %xmm3, %xmm4
    cmp      %r9d, %eax
    jle      L1
    mulq     %r8
    jmp      loop
\end{lstlisting}

\begin{lstlisting}[language=x86asm, caption={Trace of executing \texttt{L1} after \texttt{loop} in Figure~\ref{fig:motivate-example}}, label={lst:motivate-example-ordering2}, captionpos=b, numbers=left, xleftmargin=5.0ex]
    vmulps   %xmm0, %xmm1, %xmm2
    vhaddps  %xmm2, %xmm2, %xmm3
    vhaddps  %xmm3, %xmm3, %xmm4
    cmp      %r9d, %eax
    jle      L1
    movl     16(%ebp), %eax
    addl     8, %ebp
    jno      loop
\end{lstlisting}

Specifically, the \texttt{vhaddps} instruction always requires execution port 5, which is also demanded by the \texttt{mulq} instruction~\cite{uopsinfo}.
This creates a dependency between those two instructions and forces the \texttt{mulq} instruction to stall until previous \texttt{vhaddps} instructions release the desired execution port.
On the other hand, in Listing~\ref{lst:motivate-example-ordering2} \texttt{movl} and \texttt{addl} do not have conflicting resource requirements with the previous instruction.
That means both instructions will be dispatched into execution no later than the previous \texttt{vhaddps} instructions and execute in parallel, thus, resulting in higher instructions-per-cycle count than Listing~\ref{lst:motivate-example-ordering1}.

As explained earlier, current static throughput prediction approaches will 
use static instruction ordering 
as a substitute for dynamic control flow, resulting in a low-accuracy prediction.
Moreover, existing approaches face severe practical limitations with regards to scalability.
In Section~\ref{sec:design} we detail our design of \coolname{} which tackles both of these challenges.

\lstdefinelanguage{riscvasm}{
    keywords = {vsetvli, vadd.vv, vsetvl},
    alsoletter={.}
}
\begin{lstlisting}[language=riscvasm, caption={Example RISC-V assembly code}, label={lst:motivate-dynamic-info-example}, captionpos=b, numbers=left, xleftmargin=5.0ex]
    vsetvli zero, a0, e8, m2, tu, mu
    vadd.vv v12, v12, v12
    vsetvl rd, rs1, rs2
    vadd.vv v12, v12, v12
\end{lstlisting}
\subsubsection*{Execution Context.~}
Listing~\ref{lst:motivate-dynamic-info-example} shows a RISC-V assembly snippet comprised of vector instructions. In RISC-V, the \texttt{VSETVL} and \texttt{VSETVLI} instructions set the vector length multiplier (LMUL) which determines the number of elements that are processed by subsequent vector instructions. Therefore, the latency of each vector instruction differs depending on the current value of LMUL.
For instance, line~1 in Listing~\ref{lst:motivate-dynamic-info-example} sets LMUL to 2 (due to the \texttt{m2} operand), which results in a cycle latency that reflects a LMUL of 2 for the \texttt{VADD} instruction in the next line; in line~3 LMUL is set to the value stored in register \texttt{rs2}, resulting in a potentially \textit{different} latency that reflects the LMUL that was just set, for \texttt{VADD} at line 4.
Due to the lack of dynamic runtime information, static approaches can not determine the concrete value of \texttt{rs2} and therefore can not accurately model the latency of the \texttt{VADD} instruction in line~4.

This example shows that while the instructions on line~2 and 4 are identical, their exact latencies are actually influenced by the environment values, namely LMUL, as well as dynamic values stored in the registers. Current throughout prediction approaches fail to provide accurate estimations for these cases and resort to a conservative upper bound latency due to the lack of dynamic information.
Some existing tools can circumvent this issue with manual annotations. For instance, LLVM~MCA allows developers to instrument their programs with special comments that contain runtime information, which MCA uses to make more accurate queries into the scheduler model. However, while these instrument comments can improve analysis, handwriting them does not scale well with large number of instructions.

\subsubsection*{Memory Aliasing.~}
Real processors reorder instructions to optimize instruction throughput.
To that end, they analyze memory dependencies between individual load and store instructions to determine a valid instruction scheduling that minimizes contention.
Static throughput estimators trying to model this behaviour are limited due to the lack of concrete memory aliasing information. For instance, Listing~\ref{lst:motivate-example-mem-alias} shows two x86\_64 instructions that access memories indexed by base registers \texttt{\%r13} and \texttt{\%r14}.
Without knowing the exact values in these base registers, it's hard to know if memory aliasing prevents the instructions from being reordered, which makes a big difference in terms of latency.
Some tools like MCA either assumes that individual memory operations never access aliasing addresses or that all memory accesses alias;
both cases resulting in lowered prediction accuracy.
\begin{lstlisting}[language=x86asm, caption={x86\_64 assembly code with memory accesses}, label={lst:motivate-example-mem-alias}, captionpos=b, numbers=left, xleftmargin=5.0ex]
    addq  7, 8(%r13)
    movq  %r9, 8(%r14)
\end{lstlisting}

\subsection{Differential Throughput Estimation}\label{sec:usecase}
Differential throughput estimation is meant to predict the performance impacts of applying certain changes (\emph{e.g.} software patches) on the target programs.
To motivate \coolname{}'s differential throughout estimating capabilities, we detail how \coolname{} can be applied to find the optimal variant of a security patch to a binary with tight timing requirements post deployment.
In our scenario, a buffer overflow vulnerability due to a missing bounds check has been identified.
To fix the vulnerability the developer needs to patch the missing check into the binary.
As detailed in Section~\ref{sec:background} the location of the patch as well as the specific assembly instructions can have a high impact on overall performance due to resource contentions on the micro architectural level.
Developers therefore typically iterate through several semantically equivalent versions of a patch in order to minimize the performance degradation.
\coolname{} provides developers with the means to quickly iterate through several versions of the target binary to estimate the resulting performance impact and triage potential bottlenecks without requiring access to the actual hardware.

To perform the throughput estimation, a developer runs the original binary as well as several patch candidates with concrete inputs through the \coolname{} pipeline and compares estimated cycle counts between versions.
If required, \coolname{} allows to restrict the analysis to specific regions of the program (see Section~\ref{sec:design-overview-challenges}).
In addition, \coolname{} provides a timeline view which details each instruction’s state transitions through the instruction pipeline (see Section~\ref{sec:impl-viewer}).
This information helps developers to triage performance degradations and guides them towards execution paths that are less sensitive to changes.

\begin{figure}
    \centering
        \includegraphics [clip, trim=2cm 4cm 2cm 4cm, width=\columnwidth] {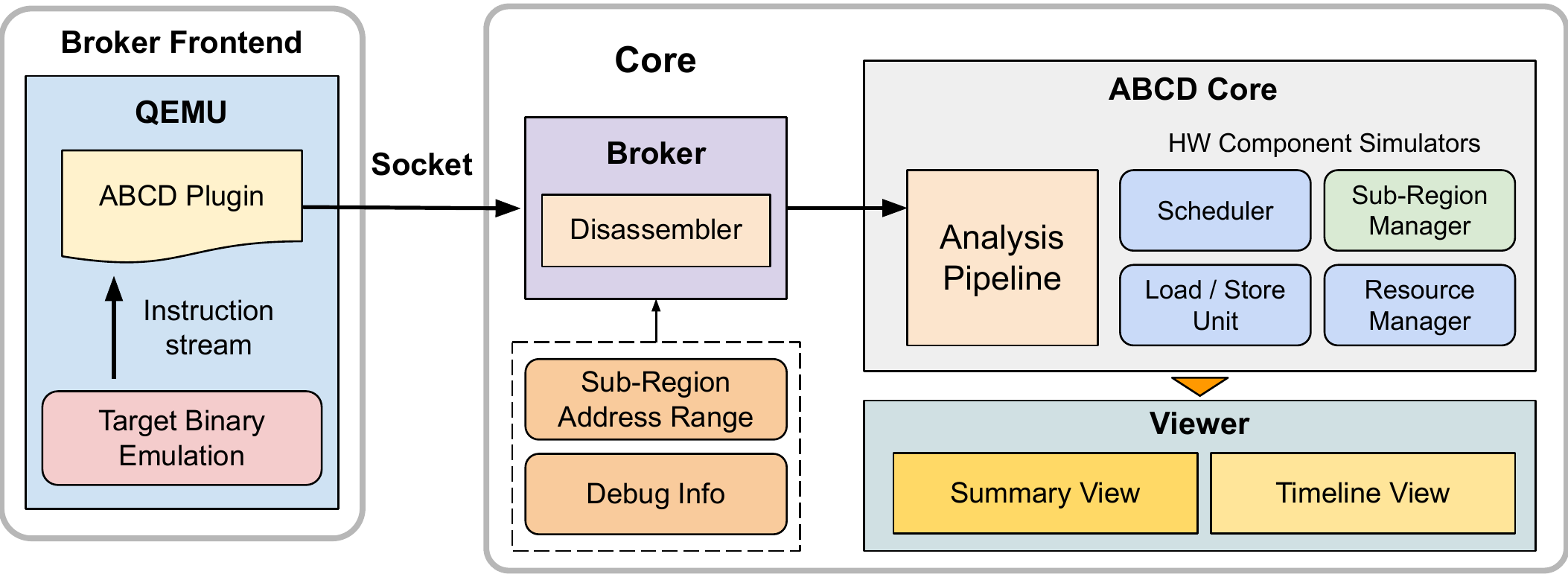}
    \caption{General workflow and major components in \coolname{}. Components with dashed outline are optional}
    \label{fig:mcad-basic-workflow}
\end{figure}
\section{Design}\label{sec:design}
In this section we present our overall design of \coolname{} depicted in Figure~\ref{fig:mcad-basic-workflow}.
As explained in the previous section, current throughput prediction approaches face severe challenges with respect to prediction across basic blocks, lack of dynamic information, as well as scalability and turnaround times.
The main goal of \coolname{} is to tackle all of these challenges to enable scalable and precise differential throughput analyses that can be used to actively drive development and steer engineers towards implementations with favorable runtime behavior.


\subsection{Goals and Challenges}\label{sec:design-overview-challenges}
\coolname{}'s design should tackle three main goals:


First, \coolname{} aims to provide whole-program throughput estimates across thousands of basic blocks and potentially millions of instructions.
At a high level, \coolname{} consists of a broker component that provides execution traces in form of instruction stream along with its dynamic contexts, and a core component that analyzes instruction-level throughput of the respective trace on-the-fly.
Results can then be processed by a viewer component for human-readable summarization and data reporting.
In principle, the method by which execution traces are obtained and streamed to the core component is not tightly coupled to the method that is used to analyze the instruction stream.
In early tests we compared several existing throughput analysis tools for use with our core component.
However, we encountered several challenges with adopting any of them for our framework.
As illustrated in the previous section, state of the art throughput prediction approaches do not generalize over dynamic contexts like control-flow transfers, register state, and memory aliasing.
Besides, as existing throughput prediction tools are designed for single basic block use, they also fail to scale up, in terms of both memory consumption and processing capabilities, when streaming input instructions on-the-fly from the broker component even for trivial programs.
We also encountered numerous bugs when using the tools that seemed most fitting in this dynamic context, some of which we detail in Section~\ref{sec:impl}.


Second, \coolname{} aims to support a development-driven workflow.
This means, that developers are able to use \coolname{} to analyze the timing impact after modifying some part of the code, which might take the form of both a binary patch or a source-level change of the original program under our model.
In addition to whole-program analyses, developers hence are able to choose to analyze only parts of the program.
Selecting which parts of the program to analyze is done at varying levels of granularity to reduce noise in the resulting reports and speed up the analysis if so required.
For example, in the scenario outlined by Section~\ref{sec:usecase}, the target program might contain components such as unmodified sequences of code, which are irrelevant to the throughput analysis, but whose run time might depend on unpredictable inputs (\emph{e.g.} random number generators). In such cases, the developer can circumvent those components by excluding their traces from \coolname{}'s analysis.

Third, \coolname{} aims to provide \emph{timely} feedback for the throughput estimates using an approach that ideally also \emph{generalizes} across architectures.
While purely dynamic approaches are already capable of producing whole-program estimates today, the associated turnaround times and costs of setting up and running full-scale system simulation can be prohibitively expensive (i.e., on the order of hours or even days)~\cite{burger1997simplescalar, yourst2007ptlsim, binkert2011gem5}.
Furthermore, existing dynamic throughput analysis tools are often tightly coupled to a specific architecture, which is why we opt for an emulation-based approach for producing execution traces inside our broker component using QEMU~\cite{bellard2005qemu}.

We will elaborate how \coolname{} tackles each of the respective challenges to achieve these goals throughout the rest of this section.



\subsection{Scalable Throughput Prediction}\label{sec:design-scalable-prediction}
As explained in Section~\ref{sec:background} all existing throughput prediction engines are designed with single-basic-block use in mind.
In our prototype we build on top of MCA~\cite{llvmmca}, a performance analysis tool and library that was designed to estimate the basic block throughput in a static fashion.
MCA employs a microarchitectural simulator to emulate an individual instruction's timeline inside an out-of-order processor.
It taps into the LLVM compiler's scheduling database, a mature and production-tested data source whose contents are curated by hardware vendors.

However, we found that MCA has difficulties to scale up in our dynamic scenario.
Like other throughput prediction engines, it does not support accurate analysis of instructions beyond a branch point or a function call out of the box.
On top of that, we found that some of the design trade-offs inside MCA make it prone to high memory pressure while processing large number of instructions.
Enabling online analysis within \coolname{} required several changes to the underlying analysis infrastructure, such as MCA's serialization, memory model, and instruction lifecycle.

In Section~\ref{sec:impl-core}, we explain our modifications of MCA used in \coolname{} to tackle the aforementioned issues and make \coolname{}'s core component scale up to real-world applications.

\subsection{Development-Driven Workflow}\label{sec:design-patch-workflow}
\coolname{} enables a development-driven workflow by providing fast whole-program throughput estimates that can easily be compared between two versions of a program.
Furthermore, for cases that only modify a small portion of the original binary, \coolname{} also provides an option to analyze only part of the binary.
In this mode, developers can designate the desired area by either specifying the symbol of a function or providing explicit address ranges in the program.
If an address range is provided, \coolname{} essentially yields the original basic-block granularity of the underlying analysis engine, while providing the flexibility of comparing the execution of multiple basic blocks at the same time.
The proposal in Section~\ref{sec:design-dynamic-info} to automatically insert runtime information via instrument comments in the execution trace lowers the burden on developers, since they no longer have to insert handwriten comments everytime the LMUL changes, nor modify existing comments when a program changes.

\subsection{Analysis Performance and Generalization Across Architectures}\label{sec:design-integration}
The broker component is responsible for supplying execution traces to the core component.
This includes interoperating with the origin of execution traces and converting them into a unified low-level representation.
This also means that the broker and core components need to work together to enable a timely and architecture-independent operation of \coolname{}.
By default, execution traces are transmitted remotely from QEMU using a custom plugin.
QEMU's emulation-based approach incurs around 30\% runtime overhead~\cite{qemuoverhead}, effectively enabling near-native execution speeds when using hardware virtualization extensions.
QEMU also has extensive support for many major architectures,\footnote{QEMU supports x86, MIPS, SPARC, ARM,  PowerPC, RISC-V among many others, including individual processor models and specific microarchitectures} meaning that \coolname{}'s broker component is able to fulfill both of these requirements.
It is noteworth to mention that \coolname{} also provides a facility for reading offline execution traces, which can be collected from executions on a physical device using any kind of tracing method available for that device.
As mentioned earlier, \coolname{}'s core component uses MCA.
Since MCA uses LLVM's infrastructure, targeting different hardware architectures and processor models in the analysis engine requires little effort.\footnote{LLVM supports x86, MIPS, SPARC, ARM, PowerPC, RISC-V among many others.}
As a result, both the broker and the core component of \coolname{} generalize well across several architectures and provide top-of-the-line performance.
Moreover, with QEMU and LLVM \coolname{} uses tools that many software developers will already be deeply familiar with.

\subsection{Supplementing Dynamic Information}\label{sec:design-dynamic-info}
As detailed in Section~\ref{sec:background}, due to unknown runtime state static throughput estimators fail to model many essential programming constructs accurately.
\coolname{} addresses this issue by providing concrete runtime information that resolves ambiguities regarding block ordering, register state and memory aliasing.

In general, the predictions of static throughput estimators to not generalize accross basic block boundaries due to the lack of information about dynamic targets and data-dependent control flows. MCA, for example, simply falls through to the next instruction upon encountering a jump or call.
\coolname{} supplements this information based on the concrete execution traces obtained from QEMU and forwarded to our analysis core.

The example illustrated by Listing~\ref{lst:motivate-dynamic-info-example} ($\S$~\ref{sec:background}) shows the difficulties of traditional throughput analysis to provide accurate predictions in the presence of run time information such as values of immediates or data in registers. Although existing tools like LLVM~MCA are able to circumvent this issue with manual annotations, such solutions don't scale well with large number of instructions common in an execution trace.
With \coolname{}, a potential solution to accurately analyze Listing~\ref{lst:motivate-dynamic-info-example} is providing necessary dynamic information of the execution trace to the analysis engine.
More specifically, the tool responsible for generating execution traces (QEMU by default) can extract LMUL when it comes across either the \texttt{VSETVL} or \texttt{VSETVLI} instruction. Such LMUL values are subsequently provided along with the trace to the analysis engine to aid precision, without any intervention from the developer.
This can easily be generalized to other applications by sending different kinds of dynamic information to the analysis engine accordingly.

MCA contains a component called Load Store Unit (LSUnit in Figure~\ref{fig:mcad-basic-workflow}) which simulates load and store reordering that could happen in the hardware scheduler of the simulated processor model.
This type of hardware optimization re-orders memory instructions to break dependencies when possible, which is largely determined by their memory aliasing properties at runtime, as shown in the example illustrated by Listing~\ref{lst:motivate-example-mem-alias} ($\S$~\ref{sec:background}).
However, without precise memory access information, MCA can only make coarse-grained assumptions, for instance, all memory instructions are aliasing with each other, which are controlled by a command line parameter.
\coolname{} leverages the memory traces collected from QEMU to improve this situation.
We modified MCA's Load Store Unit such that aliasing properties are now dictated by fine-grained memory accessing traces as provided by our QEMU plugin.
This enables our custom core component to simulate load and store reordering with higher accuracy by using dynamic information as it becomes available during execution.

\subsection{Model Assumptions}\label{sec:design-model-assumptions}
\coolname{} is able to model out-of-order and superscalar execution commonly seen in modern processors. Support for Simultaneous Multithreading (SMT), data or instruction cache and branch predictor simulation, are planned as future additions to the  project. In addition, \coolname{} currently only supports analysis of single threaded executions.

\section{Implementation}\label{sec:impl}
In this section we describe the workflow and implementation of \coolname{} in detail.
Figure~\ref{fig:mcad-basic-workflow} depicts the interaction between the different components: first, the target binary program is executed by QEMU, which collects execution traces and sends them to our analysis engine in real time.
Inside our analysis engine the executed instructions are further processed by timing analyses built on top of MCA~\cite{llvmmca}, which provides algorithms for microarchitectural simulation and instruction scheduling of modern processors.
Finally, \coolname{} provides estimates for key timing and performance metrics like the prospective cycle counts, instruction-level throughput, and the ability to identify potential bottlenecks. 

\subsection{Instruction Broker}\label{sec:impl-broker}
\coolname{}'s broker implementation is a standalone process that produces \texttt{MCInst}~\cite{introllvmmc} objects, an internal representation for machine code instructions used within LLVM, and forwards them in batches to the core component.
The broker interface is designed to be extensible and allow integration of custom implementations and enable streaming of instruction sequences to the core component from a variety of different sources.
So far, we integrated and tested two broker implementations: an assembly file broker that takes its input from an assembly file on disk and a QEMU broker that uses a QEMU plugin to communicate with the QEMU-broker process using TCP sockets to process the raw execution trace in real time before streaming them into the core component for analysis.


Broker implementations can choose to attach arbitrary metadata to the streamed instruction trace:
for instance, by attaching load and store addresses and the size of memory operations we can enable more precise dependency detection between memory accesses in the analysis engine of the core component.
In particular, \coolname{}'s QEMU plugin collects raw instructions as executed by the emulator alongside additional information regarding memory operations, which is then used to improve analysis results with respect to instruction reordering.
In this mode of operation \coolname{}'s broker dynamically instruments memory read and write operations to gather target address and size of the data.
The QEMU plugin will then send these data to the receiving core component that runs in parallel in a separate process.
Inside the core component this metadata that is attached to memory operations is then inserted into a registry that is used by the core component for joint analysis.


Developing custom brokers is straightforward and only requires implementing a few callback functions before loading them as shared libraries during runtime.
This allows users to rapidly switch between different workloads and environments depending on their needs.
It is important to note that we do not make any assumptions about a broker's internal execution model -- so long as the broker adheres to the streaming interface to supply the next batch of instructions.


\subsection{Analysis Core}\label{sec:impl-core}
Our core component builds on top of state-of-the-art throughput prediction engines, which are designed as offline tools for static throughput estimation of small sequences of machine instructions (usually at the basic block granularity).
Given a (short) sequence of assembly instructions, they provide throughput estimation results on the microarchitectural level either through end-to-end trained models for a given architecture or through simulation of the different stages inside a modern processor with varying levels of detail and manually tuned key parameters per architecture.
Unfortunately, all existing throughput prediction tools failed to scale up with our dynamic model of execution:
as an example, using the standard video and audio encoder FFmpeg~\cite{ffmpeg} executes around 20 million instructions on a Linux x86\_64 machine while decoding a short MPEG-4 video with duration of 2 seconds.
Within \coolname{}, this type of application would be considered a lightweight real-world workload.
We found that none of the existing approaches were able to analyze anywhere near this kind of workload.

However, since MCA already has support for slightly larger pieces of code compared to all other related approaches through their \emph{loop kernel} analysis, we implemented \coolname{}'s default analysis engine on top of that.
Our investigation into adopting MCA for our core component showed several failure cases while handling larger workloads.
Internally, MCA models four distinct execution stages \emph{Entry}, \emph{Dispatch}, \emph{Execute}, and \emph{Retire}.
Under \coolname{}'s workflow the instruction stream provided by the broker enters from the Entry stage and is processed by each subsequent stage sequentially.
MCA then assigns an internal data structure to each instruction to keep track of its scheduling status within the simulation pipeline.
Originally, this pipeline reads all input instructions ahead of time before the start of the analysis. Because MCA assumes those instructions to come from a file.
This property, while it aligns with the overall design goal of MCA to provide throughput estimates for only small sequences of assembler instructions, is not suited for whole program analysis.
In some of our tests, MCA consistently drained all available physical memory on the machine running the analysis due to the allocation of this internal instruction representation.

To address the scalability issue we created a new incremental mode for the MCA simulation pipeline.
In this mode, the simulation pipeline fetches input instructions incrementally.
If there are no instructions available from the input source, the pipeline will save its current state and exit.
Upon the arrival of new input instructions, the simulation will be restored and proceed from its previous state.
To reduce memory consumption, we implemented a new instruction recycling mechanism for the MCA simulation pipeline.
This instruction recycler will reclaim and collect internal instruction data structures from retired instructions, instead of releasing their memory.
These recycled data structures will then be reused to model new incoming instructions.
Our experiments showed that with this recycling mechanism, \coolname{}'s core analysis uses one third of memory on average than the unmodified MCA implementation.

MCA simulates different execution units which process individual instructions and determine results of the operation in question.
These estimates include cycle counts, potential pipeline stalling, and predictions of possible instruction re-ordering.
For instance, MCA's Load-Store Unit tracks the availability of memory operations and their (data) dependencies.
This is crucial for simulating out-of-order scheduling in modern processors, which frequently reorder memory operations based on their dependencies.
We significantly extended these existing capabilities by providing an \emph{online} analysis workflow:
by sequentially processing the incoming instruction stream and its dynamic contexts according to the simulated pipeline, \coolname{} is able to present an estimate of how arbitrarily long instruction sequences might be scheduled within the processor.

\subsection{Sub-Region Feature and Viewer~Component}\label{sec:impl-viewer}
\coolname{}'s viewer component displays throughput estimations with information like total cycle counts or potential pipeline stalls.
An example of this can be seen in Listing~\ref{lst:design-summary-view}.
We also prototyped a view of the timing itinerary of individual instruction in a timeline view.
For instance, Listing~\ref{lst:design-timeline-view} shows the timeline of the execution trace in Listing~\ref{lst:motivate-example-ordering1}. From this timeline we can easily spot the resource contention between \texttt{mulq} and \texttt{vhaddps} as mentioned in Section~\ref{sec:background}.
This particular view is a fork from the timeline view that exists in MCA.
However, the timeline view in MCA has limitations on the maximum number of analyzed instructions and cannot inspect the itinerary of a subset of an analyzed execution trace which we implemented as part of \coolname{}'s subregion feature.
For this reason, third-party visualization tools are also supported in this component.
For instance, we prototyped a new timeline view based on Chrome Developer Tools~(DevTools)~\cite{chromedevtool}.
In our early test we already found this a lot easier to scroll and navigate through the thousands or even millions of instructions that are processed by \coolname{}, compared to the terminal-based LLVM~MCA timeline view.
Since this view was designed to analyze large number of network requests it provides a solid basis to help our new timeline view scale up.

\begin{lstlisting}[caption={Summary View}, label={lst:design-summary-view}, captionpos=b, basicstyle=\ttfamily]
Instructions:      350
Total Cycles:      262
Total uOps:        600
Dispatch Width:    6
uOps Per Cycle:    2.29
IPC:               1.34
Block RThroughput: 5.0
\end{lstlisting}

\begin{lstlisting}[caption={Timeline View}, label={lst:design-timeline-view}, captionpos=b, basicstyle=\ttfamily]
[1,0]     . D=eeeeE---------R ..   vmulps
[1,1]     .  D====eeeeeeE---R ..   vhaddps
[1,2]     .  D==========eeeeeeER   vhaddps
[1,3]     .   D==eE------------R   cmpl
[1,4]     .   D===eE-----------R   jle
[1,5]     .   D=====eeeeE------R   mulq
[1,6]     .   DeE--------------R   jmp
\end{lstlisting}

\section{Evaluation}\label{sec:eval}
\coolname{}'s main goal is to enable developers to quickly assess and iterate on the timing impact of program modifications, including \emph{patches}, across control transfers (\emph{e.g.} branches and function calls). In this section, we use binary programs of different release versions to evaluate performance and cycle-count accuracy against physical hardware traces to quantify how \coolname{} fares in comparison.
More formally, given two different versions $i$ and $j$ of a program~$P$, denoted as $P_i$ and $P_j$, as well as a throughput predictor $H$ that provides the number of execution cycles under a specific input for the respective program, we define the \emph{differential throughput}~$\Delta_H(P_i,P_j)$ describing the change in cycle counts between version $i$ and version $j$ of program $P$ as predicted by $H$~as~follows:
$$
\Delta_H(P_i, P_j) = \frac{H(P_j)}{H(P_i)} 
$$
Given two versions of a program $P_i$ and $P_j$, as well as their inputs, we first use \coolname{} to predict their differential throughput, resulting in $\Delta_{\coolname{}}(P_i, P_j)$.
Second, we similarly measure their relative difference in cycle counts from version $i$ to version $j$ using hardware-performance counters on physical devices, resulting in ground truth differential throughput $\Delta_G(P_i, P_j)$.
Finally, we formally define the error of \coolname{}'s prediction of differential throughput between version $i$ and version $j$ as:
$$
E_{ij} = \lvert \Delta_G(P_i, P_j) - \Delta_{\coolname{}}(P_i, P_j) \rvert
$$

\subsection{Differential Throughput Prediction}\label{sec:eval-accuracy}
To assess the overall accuracy and ability to generalize throughput predictions across control transfers we conduct experiments using two popular and widely used applications as target programs: the \texttt{ffmpeg} video encoder/decoder and the C/C++/Objective-C compiler \texttt{clang}.
We select \texttt{ffmpeg} for our case study as video encoding represents a complex and highly performance-intensive task with many applications in real-world use cases.
\texttt{clang} represents a large-scale software consisting of complex branching logic, which is well suited to test \coolname{}'s cross-branch prediction accuracy and also plays an important role in many real-world scenarios.

We collect baseline cycle-count measurements on three physical machines with different instruction set architectures (ISAs) and microarchitectures:
\begin{itemize}
  \item \emph{Intel CoffeeLake:} 6-core Intel i7~8700K x86\_64 CPU, clocked at 3.70GHz and 32G of RAM running Ubuntu 20.04
  \item \emph{AMD Zen 2:} 12-core AMD Ryzen~9 3900X x86\_64 CPU, clocked at 2.48GHz and 32G of RAM running Ubuntu 20.04
  \item \emph{ARM Cortex-A57:} 4-core ARM Cortex-A57 AArch64 CPU, clocked at 1.73GHz and 4G of RAM running Ubuntu 16.04
\end{itemize}

Compared to running baseline measurements on smaller program scopes (\emph{e.g.} a single basic block), measuring larger execution traces faces much more operating system noise. To avoid noise due to CPU migration or context switching we allocate a single processor core exclusively for the process under measurement. In addition, we disable Simultaneous Multithreading (SMT, also called \emph{Hyper-Threading} on Intel processors) on the benchmarking core since it is not supported by our analysis engine as mentioned in Section~\ref{sec:design-model-assumptions}. The baseline measurements are obtained using Linux Perf, which leverages the Performance Monitor Unit (PMU) provided by the underlying hardware, and are averaged over 1000 repetitions.

\begin{figure*}
  \begin{tikzpicture}
  \begin{axis}[height=1.6cm, hide axis, xmin=0, xmax=0, ymin=0,ymax=0, legend style={font=\scriptsize,legend cell align=left, draw=none},legend image post style={scale=0.5}, mark=diamond*, mark size=1.3pt, mark options=solid]
    \addlegendimage{thick, mcadDTPPerf}
    \addlegendentry{Perf Diff. Throughput};
  \end{axis}
\end{tikzpicture}
\begin{tikzpicture}
  \begin{axis}[height=1.6cm, hide axis, xmin=0, xmax=0, ymin=0,ymax=0, legend style={font=\scriptsize,legend cell align=left, draw=none},legend image post style={scale=0.5}, mark=o, mark size=1.3pt, mark options=solid]
    \addlegendimage{thick, mcadDTPMcad}
    \addlegendentry{\coolname{} Diff. Throughput};
  \end{axis}
\end{tikzpicture}
\begin{tikzpicture}
  \begin{axis}[height=1.6cm, hide axis, xmin=0, xmax=0, ymin=0,ymax=0, legend style={font=\scriptsize,legend cell align=left, draw=none},legend image post style={scale=0.5}]
    \addlegendimage{thick, mcadGMeanError}
    \addlegendentry{Geomean Error};
  \end{axis}
\end{tikzpicture}
\begin{tikzpicture}
  \begin{axis}[height=1.6cm, hide axis, xmin=0, xmax=0, ymin=0,ymax=0, area legend, legend style={font=\scriptsize,legend cell align=left, draw=none},legend image post style={scale=0.5}]
      \addlegendimage{black, fill=mcadError}
      \addlegendentry{Error};
  \end{axis}
\end{tikzpicture}
  \vspace{-2ex}
  \centering
  \subfloat[\texttt{ffmpeg} Intel~Coffee~Lake]{
  \pgfplotstableread[
   col sep=comma,
]{./dtp_ffmpeg_coffeelake.csv}\coffeelaketable
\pgfplotsset{
  every tick label/.append style={font=\scriptsize},
}
\begin{tikzpicture}
\begin{axis}[
      ybar,
      yticklabels={,,},
      xticklabels from table={\coffeelaketable}{Ver},
      x tick label style={rotate=25,font=\scriptsize},
      xtick={0,1,2,3,4,5,6,7},
      xtick pos=left,
      ytick pos=right,
      ymajorgrids=true,
      ymax=16,
      ymin=0,
      width=6.4cm,
      height=3.0cm,
   ]
   \addplot[black, fill=mcadError] table [
      col sep=comma,
      y=Error,
      x expr=\coordindex,
   ]{\coffeelaketable};
\end{axis}
\begin{axis}[
     xtick={0,1,2,3,4,5,6,7},
     yticklabels={,,},
     xticklabels={,,},
     draw=none,
     ytick style={draw=none},
     xtick style={draw=none},
     width=6.4cm,
     height=3.0cm,
     ymin=0,
     ymax=16,
 ]
 \addplot [mcadGMeanError, thick] table [
     col sep=comma,
     x expr=\coordindex,
     y=GMean,
 ]{\coffeelaketable};
\end{axis}
\begin{axis}[
        ylabel near ticks,
        y tick label style={font=\scriptsize},
        y label style={font=\scriptsize},
        ylabel = {Diff. Throughput},
        xticklabels={,,},
        xtick pos=left,
        width=6.4cm,
        height=3.0cm,
        ymin=0,
        ymax=1.6,
    ]
    \addplot [mcadDTPPerf, thick, mark=diamond*, mark size=1.3pt, mark options=solid] table [
        col sep=comma,
        x expr=\coordindex,
        y=CoffeeLake,
    ]{\coffeelaketable};
    \addplot [mcadDTPMcad, thick, mark=o, mark size=1.3pt, mark options=solid] table [
        col sep=comma,
        x expr=\coordindex,
        y=MCAD,
    ]{\coffeelaketable};
\end{axis}
\end{tikzpicture}
  \label{fig:ffmpeg-skl}
  }
  \subfloat[\texttt{ffmpeg} AMD~Zen2]{
  \pgfplotstableread[
   col sep=comma,
]{./dtp_ffmpeg_zen.csv}\zentable
\pgfplotsset{
  every tick label/.append style={font=\scriptsize},
}
\begin{tikzpicture}
\begin{axis}[
      ybar,
      yticklabels={,,},
      xticklabels from table={\zentable}{Ver},
      x tick label style={rotate=25,font=\scriptsize},
      xtick={0,1,2,3,4,5,6,7},
      xtick pos=left,
      ytick pos=right,
      ymajorgrids=true,
      ymax=16,
      ymin=0,
      width=6.4cm,
      height=3.0cm,
   ]
   \addplot[black, fill=mcadError] table [
      col sep=comma,
      y=Error,
      x expr=\coordindex,
   ]{\zentable};
\end{axis}
\begin{axis}[
     xtick={0,1,2,3,4,5,6,7},
     yticklabels={,,},
     xticklabels={,,},
     draw=none,
     ytick style={draw=none},
     xtick style={draw=none},
     width=6.4cm,
     height=3.0cm,
     ymin=0,
     ymax=16,
 ]
 \addplot [mcadGMeanError, thick] table [
     col sep=comma,
     x expr=\coordindex,
     y=GMean,
 ]{\zentable};
\end{axis}
\begin{axis}[
        y tick label style={font=\scriptsize},
        xticklabels={,,},
        yticklabels={,,},
        xtick={0,1,2,3,4,5,6,7},
        xtick pos=left,
        width=6.4cm,
        height=3.0cm,
        yticklabel pos=right,
        ymin=0,
        ymax=1.6,
    ]
    \addplot [mcadDTPPerf, thick, mark=diamond*, mark size=1.3pt, mark options=solid] table [
        col sep=comma,
        x expr=\coordindex,
        y=Zen2,
    ]{\zentable};
    \addplot [mcadDTPMcad, thick, mark=o, mark size=1.3pt, mark options=solid] table [
        col sep=comma,
        x expr=\coordindex,
        y=MCAD,
    ]{\zentable};
\end{axis}
\end{tikzpicture}
  \label{fig:ffmpeg-zen2}
  }
  \subfloat[\texttt{ffmpeg} ARM~Cortex-A57]{
  \pgfplotstableread[
   col sep=comma,
]{./dtp_ffmpeg_arm.csv}\armtable
\pgfplotsset{
  every tick label/.append style={font=\scriptsize},
}
\begin{tikzpicture}
\begin{axis}[
      ybar,
      ylabel near ticks,
      y label style={font=\scriptsize},
      ylabel = {Error (Percentage)},
      xticklabels from table={\armtable}{Ver},
      x tick label style={rotate=25,font=\scriptsize},
      xtick={0,1,2,3,4,5,6,7},
      xtick pos=left,
      ytick pos=right,
      ymajorgrids=true,
      ymax=16,
      ymin=0,
      width=6.4cm,
      height=3.0cm,
   ]
   \addplot[black, fill=mcadError] table [
      col sep=comma,
      y=Error,
      x expr=\coordindex,
   ]{\armtable};
\end{axis}
\begin{axis}[
     xtick={0,1,2,3,4,5,6,7},
     yticklabels={,,},
     xticklabels={,,},
     draw=none,
     ytick style={draw=none},
     xtick style={draw=none},
     width=6.4cm,
     height=3.0cm,
     ymin=0,
     ymax=16,
 ]
 \addplot [mcadGMeanError, thick] table [
     col sep=comma,
     x expr=\coordindex,
     y=GMean,
 ]{\armtable};
\end{axis}
\begin{axis}[
        y tick label style={font=\scriptsize},
        xticklabels={,,},
        yticklabels={,,},
        xtick={0,1,2,3,4,5,6,7},
        xtick pos=left,
        width=6.4cm,
        height=3.0cm,
        yticklabel pos=right,
        ymin=0,
        ymax=1.6,
    ]
    \addplot [mcadDTPPerf, thick, mark=diamond*, mark size=1.3pt, mark options=solid] table [
        col sep=comma,
        x expr=\coordindex,
        y=Cortex-A57,
    ]{\armtable};
    \addplot [mcadDTPMcad, thick, mark=o, mark size=1.3pt, mark options=solid] table [
        col sep=comma,
        x expr=\coordindex,
        y=MCAD,
    ]{\armtable};
\end{axis}
\end{tikzpicture}
  \label{fig:ffmpeg-cortex}
  }

  \subfloat[\texttt{Clang} Intel~Coffee~Lake]{
  \pgfplotstableread[
   col sep=comma,
]{./dtp_clang_coffeelake.csv}\coffeelaketable
\pgfplotsset{
  every tick label/.append style={font=\scriptsize},
}
\begin{tikzpicture}
\begin{axis}[
      ybar,
      yticklabels={,,},
      xticklabels from table={\coffeelaketable}{Ver},
      x tick label style={rotate=25,font=\scriptsize},
      xtick={0,1,2,3,4,5,6,7},
      xtick pos=left,
      ytick pos=right,
      ymajorgrids=true,
      ymax=16,
      ymin=0,
      width=6.4cm,
      height=3.0cm,
   ]
   \addplot[black, fill=mcadError] table [
      col sep=comma,
      y=Error,
      x expr=\coordindex,
   ]{\coffeelaketable};
\end{axis}
\begin{axis}[
     xtick={0,1,2,3,4,5,6,7},
     yticklabels={,,},
     xticklabels={,,},
     draw=none,
     ytick style={draw=none},
     xtick style={draw=none},
     width=6.4cm,
     height=3.0cm,
     ymin=0,
     ymax=16,
 ]
 \addplot [mcadGMeanError, thick] table [
     col sep=comma,
     x expr=\coordindex,
     y=GMean,
 ]{\coffeelaketable};
\end{axis}
\begin{axis}[
        ylabel near ticks,
        y tick label style={font=\scriptsize},
        y label style={font=\scriptsize},
        ylabel = {Diff. Throughput},
        xticklabels={,,},
        xtick pos=left,
        width=6.4cm,
        height=3.0cm,
        ymin=0,
        ymax=1.6,
    ]
    \addplot [mcadDTPPerf, thick, mark=diamond*, mark size=1.3pt, mark options=solid] table [
        col sep=comma,
        x expr=\coordindex,
        y=CoffeeLake,
    ]{\coffeelaketable};
    \addplot [mcadDTPMcad, thick, mark=o, mark size=1.3pt, mark options=solid] table [
        col sep=comma,
        x expr=\coordindex,
        y=MCAD,
    ]{\coffeelaketable};
\end{axis}
\end{tikzpicture}
  \label{fig:clang-skl}
  }
  \subfloat[\texttt{Clang} AMD~Zen2]{
  \pgfplotstableread[
   col sep=comma,
]{./dtp_clang_zen.csv}\zentable
\pgfplotsset{
  every tick label/.append style={font=\scriptsize},
}
\begin{tikzpicture}
\begin{axis}[
      ybar,
      yticklabels={,,},
      xticklabels from table={\zentable}{Ver},
      x tick label style={rotate=25,font=\scriptsize},
      xtick={0,1,2,3,4,5,6,7},
      xtick pos=left,
      ytick pos=right,
      ymajorgrids=true,
      ymax=16,
      ymin=0,
      width=6.4cm,
      height=3.0cm,
   ]
   \addplot[black, fill=mcadError] table [
      col sep=comma,
      y=Error,
      x expr=\coordindex,
   ]{\zentable};
\end{axis}
\begin{axis}[
     xtick={0,1,2,3,4,5,6,7},
     yticklabels={,,},
     xticklabels={,,},
     draw=none,
     ytick style={draw=none},
     xtick style={draw=none},
     width=6.4cm,
     height=3.0cm,
     ymin=0,
     ymax=16,
 ]
 \addplot [mcadGMeanError, thick] table [
     col sep=comma,
     x expr=\coordindex,
     y=GMean,
 ]{\zentable};
\end{axis}
\begin{axis}[
        y tick label style={font=\scriptsize},
        xticklabels={,,},
        yticklabels={,,},
        xtick={0,1,2,3,4,5,6,7},
        xtick pos=left,
        width=6.4cm,
        height=3.0cm,
        yticklabel pos=right,
        ymin=0,
        ymax=1.6,
    ]
    \addplot [mcadDTPPerf, thick, mark=diamond*, mark size=1.3pt, mark options=solid] table [
        col sep=comma,
        x expr=\coordindex,
        y=Zen2,
    ]{\zentable};
    \addplot [mcadDTPMcad, thick, mark=o, mark size=1.3pt, mark options=solid] table [
        col sep=comma,
        x expr=\coordindex,
        y=MCAD,
    ]{\zentable};
\end{axis}
\end{tikzpicture}
  \label{fig:clang-zen2}
  }
  \subfloat[\texttt{Clang} ARM~Cortex-A57]{
  \pgfplotstableread[
   col sep=comma,
]{./dtp_clang_arm.csv}\armtable
\pgfplotsset{
  every tick label/.append style={font=\scriptsize},
}
\begin{tikzpicture}
\begin{axis}[
      ybar,
      ylabel near ticks,
      y label style={font=\scriptsize},
      ylabel = {Error (Percentage)},
      xticklabels from table={\armtable}{Ver},
      x tick label style={rotate=25,font=\scriptsize},
      xtick={0,1,2,3,4,5,6,7},
      xtick pos=left,
      ytick pos=right,
      ymajorgrids=true,
      ymax=16,
      ymin=0,
      width=6.4cm,
      height=3.0cm,
   ]
   \addplot[black, fill=mcadError] table [
      col sep=comma,
      y=Error,
      x expr=\coordindex,
   ]{\armtable};
\end{axis}
\begin{axis}[
     xtick={0,1,2,3,4,5,6,7},
     yticklabels={,,},
     xticklabels={,,},
     draw=none,
     ytick style={draw=none},
     xtick style={draw=none},
     width=6.4cm,
     height=3.0cm,
     ymin=0,
     ymax=16,
 ]
 \addplot [mcadGMeanError, thick] table [
     col sep=comma,
     x expr=\coordindex,
     y=GMean,
 ]{\armtable};
\end{axis}
\begin{axis}[
        y tick label style={font=\scriptsize},
        xticklabels={,,},
        yticklabels={,,},
        xtick={0,1,2,3,4,5,6,7},
        xtick pos=left,
        width=6.4cm,
        height=3.0cm,
        yticklabel pos=right,
        ymin=0,
        ymax=1.6,
    ]
    \addplot [mcadDTPPerf, thick, mark=diamond*, mark size=1.3pt, mark options=solid] table [
        col sep=comma,
        x expr=\coordindex,
        y=Cortex-A57,
    ]{\armtable};
    \addplot [mcadDTPMcad, thick, mark=o, mark size=1.3pt, mark options=solid] table [
        col sep=comma,
        x expr=\coordindex,
        y=MCAD,
    ]{\armtable};
\end{axis}
\end{tikzpicture}
  \label{fig:clang-cortex}
  }
%
%
%
  \caption{Differential execution timing comparisons between subsequent development versions of \texttt{ffmpeg} and \texttt{clang}. Relative cycle count differences are plotted in blue and orange. Gray bars represent MAPE with geo. mean plotted in red.}
  \vspace*{-1.5em}
\label{fig:clang-main-eval}
\end{figure*}
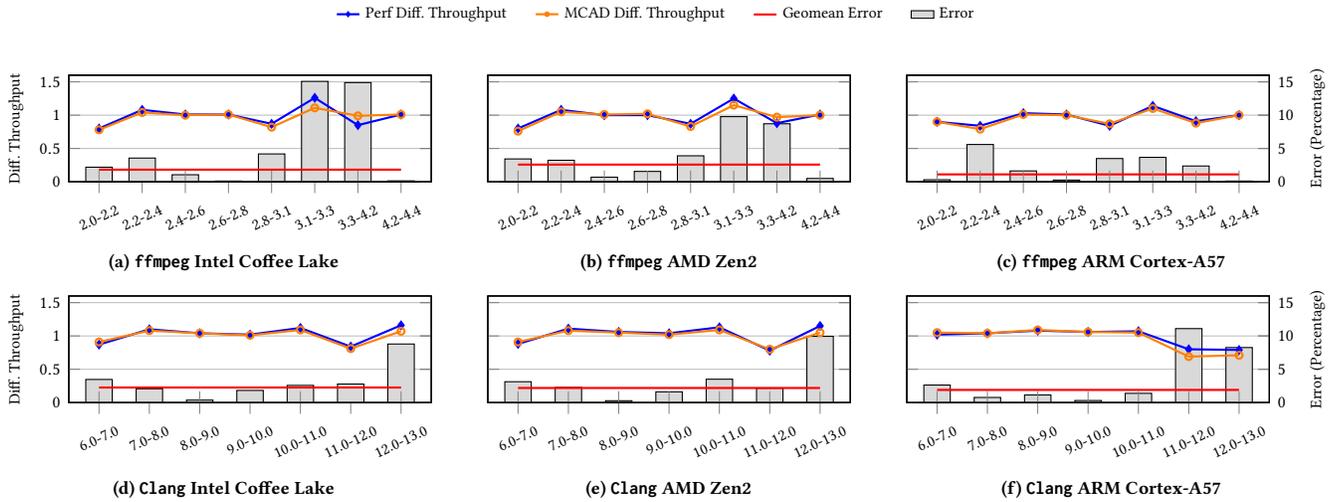

\subsubsection{FFmpeg}\label{sec:eval-ffmpeg}
We evaluate \texttt{ffmpeg} on subsequent version pairs using 9 different release versions in the following order: 2.0, 2.2, 2.4, 2.6, 2.8, 3.1, 3.3, 4.2, and 4.4.
For each experiment, we use the same 14KB MPEG-4 video file as reference input and execute the following command:
\begin{lstlisting}[basicstyle=\ttfamily]
  ffmpeg -i input.mp4 -f null -
\end{lstlisting}
Figure~\ref{fig:ffmpeg-skl}, \ref{fig:ffmpeg-zen2}, and \ref{fig:ffmpeg-cortex} depict the differential throughput predictions of \coolname{}
$\Delta_{\coolname{}}(\mathtt{ffmpeg}_i,\ \mathtt{ffmpeg}_j)$
and the baseline $\Delta_{ij}$ for version pairs $(i,j)$ as $(2.0, 2.2)$, $(2.2, 2.4)$, and so forth.
In addition, the Figures present the resulting error rates $E_{ij}$ for the corresponding version pairs between ground truth measurements on physical devices and predictions by \coolname{}.

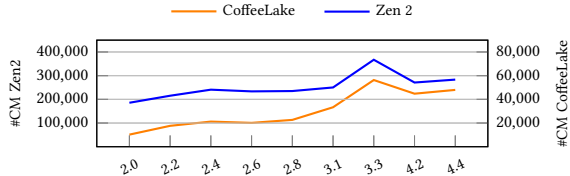
\begin{figure}
    \begin{tikzpicture}
      \begin{axis}[height=1.6cm, hide axis, xmin=0, xmax=0, ymin=0,ymax=0, legend style={font=\scriptsize,legend cell align=left, draw=none}]
        \addlegendimage{thick, orange}
        \addlegendentry{CoffeeLake};
      \end{axis}
    \end{tikzpicture}
    \begin{tikzpicture}
      \begin{axis}[height=1.6cm, hide axis, xmin=0, xmax=0, ymin=0,ymax=0, legend style={font=\scriptsize,legend cell align=left, draw=none}]
        \addlegendimage{thick, blue}
        \addlegendentry{Zen 2};
      \end{axis}
    \end{tikzpicture}

  \centering
  \pgfplotstableread[
   col sep=comma,
]{./cache_ffmpeg.csv}\cachetable
\pgfplotsset{scaled y ticks=false}
\pgfplotsset{
  every tick label/.append style={font=\scriptsize},
}
\begin{tikzpicture}
\begin{axis}[
        ylabel near ticks,
        y label style={font=\scriptsize},
        ylabel = {\#CM CoffeeLake},
        yticklabel style={
           /pgf/number format/fixed,
           /pgf/number format/precision=5
        },
        scaled y ticks=false,
        y tick label style={font=\scriptsize},
        ytick={20000,40000,60000,80000},
        x tick label style={rotate=25,font=\scriptsize},
        xticklabels from table={\cachetable}{Ver},
        xtick={0,1,2,3,4,5,6,7,8},
        xtick pos=left,
        width=0.8\columnwidth,
        height=3.0cm,
        yticklabel pos=right,
        ymajorgrids=true,
        ymin=0,
        ymax=90000,
    ]
    \addplot [orange, thick] table [
        col sep=comma,
        x expr=\coordindex,
        y=CoffeeLake,
    ]{\cachetable};
\end{axis}
\begin{axis}[
        ylabel near ticks,
        y label style={font=\scriptsize},
        ylabel = {\#CM Zen2},
        yticklabel style={
           /pgf/number format/fixed,
           /pgf/number format/precision=5
        },
        scaled y ticks=false,
        y tick label style={font=\scriptsize},
        ytick={100000,200000,300000,400000},
        x tick label style={rotate=25,font=\scriptsize},
        xticklabels from table={\cachetable}{Ver},
        xtick={0,1,2,3,4,5,6,7,8},
        xticklabels={,,},
        xtick pos=left,
        width=0.8\columnwidth,
        height=3.0cm,
        yticklabel pos=left,
        ymajorgrids=true,
        ymin=0,
        ymax=450000,
    ]
    \addplot [blue, thick] table [
        col sep=comma,
        x expr=\coordindex,
        y=Zen2,
    ]{\cachetable};
\end{axis}
\end{tikzpicture}
  \vspace*{-0.7em}
  \caption{Number of cache misses during the execution of different \texttt{ffmpeg} versions on CoffeeLake and Zen2 machines.}
  \label{fig:ffmpeg-x86-cache-misses}
\end{figure}

Our results show that \coolname{} closely follows hardware cycle counts and never deviates from changes in the baseline count by more than 15\%.
On average, the error is \ffmpegIntelError{} for Intel, \ffmpegAmdError{} for AMD, and \ffmpegArmError{} for the ARM Cortex-A57 with a standard deviation of less than \ffmpegErrorStd{} in all cases.
Nevertheless, on both Intel and AMD machines, two version pairs show unusually high error: $(3.1, 3.3)$ and $(3.3, 4.2)$ deviate from baseline measurements by around 10\% to 15\%. Further investigation showed that this deviation is likely due to a high number of cache misses during the execution of \texttt{ffmpeg} version 3.3 on Intel and AMD machines.
Figure~\ref{fig:ffmpeg-x86-cache-misses} shows the number of cache misses when running different versions of \texttt{ffmpeg} on CoffeeLake and Zen~2, measured using hardware performance counters averaged over 1000 repetitions. On both machines the number of cache misses spikes for version 3.3, exceeding the second highest measurement by at most 30\%. As detailed in Section~\ref{sec:design-scalable-prediction}, \coolname{} utilizes LLVM's instruction scheduling database for instruction latency information which assumes that all memory accesses result in cache hits. Therefore, without proper and potentially expensive cache simulation (See Section~\ref{sec:design-model-assumptions}), \coolname{}'s precision will be hindered by cycle count penalties originating from cache misses.

\subsubsection{Clang}\label{sec:eval-clang}
We conducted our experiments for \texttt{clang} using 8 different release versions as follows: 6.0, 7.0, 8.0, 9.0, 10.0, 11.0, 12.0, and 13.0.
In order to reduce the amount of I/O interference and simplify the experiments without losing generality, we focus on the backend of \texttt{clang}'s compilation pipeline. More specifically, we measure the cycles consumed by \texttt{clang} in compiling an unoptimized LLVM IR program to an object file.
The LLVM IR input file (\texttt{input.ll}) is generated from the following C program:
\begin{lstlisting}[basicstyle=\ttfamily]
int foo(int x, int y) {
  return x * 2 + y;
}
\end{lstlisting}
We use \texttt{input.ll} as the reference input for our experiments and execute the following command:
\begin{lstlisting}[basicstyle=\ttfamily]
clang -O2 -c input.ll -o /dev/null
\end{lstlisting}
Figure~\ref{fig:clang-skl}, ~\ref{fig:clang-zen2}, and ~\ref{fig:clang-cortex} depict the differential throughput predictions of \coolname{} $\Delta_{\coolname{}}(\mathtt{clang}_i,\ \mathtt{clang}_j)$ and the baseline $\Delta_{ij}$ as well as the error rates $E_{ij}$ between predictions by \coolname{} and baseline for version pairs $(i,j)$ as $(6, 7)$, $(7, 8)$, and so forth.

Again, our results show that \coolname{} closely follows the hardware cycle count and never deviates from the changes in the baseline count by more than 12\%.
On average, the error is \clangIntelError{} for Intel, \clangAmdError{} for AMD, and \clangArmError{} for the ARM Cortex-A57 with a standard deviation of less than \clangErrorStd{} in all cases.
On both Intel and AMD machines, we observe a spike of nearly 10\% error on version pair $(12.0, 13.0)$. Similar to the culprit for unusually high error percentage in Section~\ref{sec:eval-ffmpeg}, we find that this spike of error is caused by higher number of cache misses on version 13.0: compared to other \texttt{clang} versions, \texttt{clang} 13.0 creates 13\% to 35\% more cache misses on both machines.
In addition, on the ARM machine, we observe around 10\% of error on version pairs $(11.0, 12.0)$ and $(12.0, 13.0)$. We find that this is caused by sudden increase of memory operations in both version pairs. For instance, the number of \texttt{LDUR} (memory load) and \texttt{STR} (memory store) instructions increases by a factor of $27$ in said version pairs.
Modern processors usually apply advanced optimizations, such as load-store forwarding, during the execution of these memory operations which might not be accurately modeled by MCA.

\subsection{Differential Throughput Prediction on Small Changes}
So far, \coolname{} has shown its accurate differential throughput predictions on changes between major software release versions. In this sub-section, we futher illustrate \coolname{}'s capability of predicting differential throughput originating from (much) smaller changes, like software patches. This is supported by repeating the Intel CoffeeLake experiments from Section~\ref{sec:eval-ffmpeg} and Section~\ref{sec:eval-clang}, but using \textit{minor} \texttt{ffmpeg} and \texttt{clang} releases, which have far less modifications between them, rather than their major releases as the analysis targets. Ideally, we should characterize the size of changes between two different versions based on their program binaries. However, measuring differences between two binaries is a hard problem~\cite{duan2020deepbindiff,ren2021unleashing}. Therefore, we use lines of (source) code changes (\#~LoCC) as an approximate metric of change between two software versions.

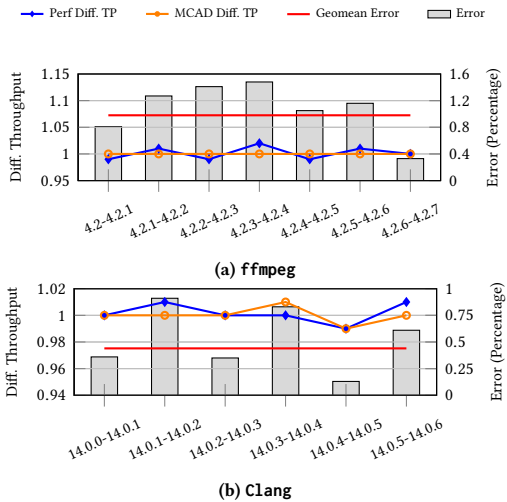
\begin{figure}
\centering
\begin{tikzpicture}
  \begin{axis}[height=1.6cm, hide axis, xmin=0, xmax=0, ymin=0,ymax=0, legend style={font=\tiny,legend cell align=left, draw=none},legend image post style={scale=0.5}, mark=diamond*, mark size=1.3pt, mark options=solid]
    \addlegendimage{thick, mcadDTPPerf}
    \addlegendentry{Perf Diff. TP};
  \end{axis}
\end{tikzpicture}
\begin{tikzpicture}
  \begin{axis}[height=1.6cm, hide axis, xmin=0, xmax=0, ymin=0,ymax=0, legend style={font=\tiny,legend cell align=left, draw=none},legend image post style={scale=0.5}, mark=o, mark size=1.3pt, mark options=solid]
    \addlegendimage{thick, mcadDTPMcad}
    \addlegendentry{\coolname{} Diff. TP};
  \end{axis}
\end{tikzpicture}
\begin{tikzpicture}
  \begin{axis}[height=1.6cm, hide axis, xmin=0, xmax=0, ymin=0,ymax=0, legend style={font=\tiny,legend cell align=left, draw=none},legend image post style={scale=0.5}]
    \addlegendimage{thick, mcadGMeanError}
    \addlegendentry{Geomean Error};
  \end{axis}
\end{tikzpicture}
\begin{tikzpicture}
  \begin{axis}[height=1.6cm, hide axis, xmin=0, xmax=0, ymin=0,ymax=0, area legend, legend style={font=\tiny,legend cell align=left, draw=none},legend image post style={scale=0.5}]
      \addlegendimage{black, fill=mcadError}
      \addlegendentry{Error};
  \end{axis}
\end{tikzpicture}

\subfloat[\texttt{ffmpeg}]{
\pgfplotstableread[
   col sep=comma,
]{./ffmpeg_minor.csv}\coffeelaketable
\pgfplotsset{
  every tick label/.append style={font=\scriptsize},
}
\begin{tikzpicture}
\begin{axis}[
      ybar,
      ylabel near ticks,
      y label style={font=\scriptsize},
      ylabel = {Error (Percentage)},
      xticklabels from table={\coffeelaketable}{Ver},
      x tick label style={rotate=25,font=\scriptsize},
      xtick={0,1,2,3,4,5,6,7},
      xtick pos=left,
      ytick={0.0, 0.4, 0.8, 1.2, 1.6},
      ytick pos=right,
      ymajorgrids=true,
      ymax=1.6,
      ymin=0,
      width=6.4cm,
      height=3.0cm,
   ]
   \addplot[black, fill=mcadError] table [
      col sep=comma,
      y=Error,
      x expr=\coordindex,
   ]{\coffeelaketable};
\end{axis}
\begin{axis}[
     xtick={0,1,2,3,4,5,6,7},
     yticklabels={,,},
     xticklabels={,,},
     draw=none,
     ytick style={draw=none},
     xtick style={draw=none},
     width=6.4cm,
     height=3.0cm,
     ymin=0,
     ymax=1.6,
 ]
 \addplot [mcadGMeanError, thick] table [
     col sep=comma,
     x expr=\coordindex,
     y=GMean,
 ]{\coffeelaketable};
\end{axis}
\begin{axis}[
        ylabel near ticks,
        y tick label style={font=\scriptsize},
        y label style={font=\scriptsize},
        ylabel = {Diff. Throughput},
        xticklabels={,,},
        xtick pos=left,
        width=6.4cm,
        height=3.0cm,
        ymajorgrids=true,
        ytick={0.95, 1.0, 1.05, 1.1, 1.15},
        ymin=0.95,
        ymax=1.15,
    ]
    \addplot [mcadDTPPerf, thick, mark=diamond*, mark size=1.3pt, mark options=solid] table [
        col sep=comma,
        x expr=\coordindex,
        y=CoffeeLake,
    ]{\coffeelaketable};
    \addplot [mcadDTPMcad, thick, mark=o, mark size=1.3pt, mark options=solid] table [
        col sep=comma,
        x expr=\coordindex,
        y=MCAD,
    ]{\coffeelaketable};
\end{axis}
\end{tikzpicture}
\label{fig:ffmpeg-small-eval}
}
\vspace*{-1.5em}
\subfloat[\texttt{Clang}]{
\pgfplotstableread[
   col sep=comma,
]{./clang_minor.csv}\coffeelaketable
\pgfplotsset{
  every tick label/.append style={font=\scriptsize},
}
\begin{tikzpicture}
\begin{axis}[
      ybar,
      ylabel near ticks,
      y label style={font=\scriptsize},
      ylabel = {Error (Percentage)},
      xticklabels from table={\coffeelaketable}{Ver},
      x tick label style={rotate=25,font=\scriptsize},
      ytick={0.0, 0.25, 0.5, 0.75, 1.0},
      xtick={0,1,2,3,4,5,6,7},
      xtick pos=left,
      ytick pos=right,
      ymajorgrids=true,
      ymax=1.0,
      ymin=0,
      width=6.4cm,
      height=3.0cm,
   ]
   \addplot[black, fill=mcadError] table [
      col sep=comma,
      y=Error,
      x expr=\coordindex,
   ]{\coffeelaketable};
\end{axis}
\begin{axis}[
     xtick={0,1,2,3,4,5,6,7},
     yticklabels={,,},
     xticklabels={,,},
     draw=none,
     ytick style={draw=none},
     xtick style={draw=none},
     width=6.4cm,
     height=3.0cm,
     ymin=0,
     ymax=1.0,
 ]
 \addplot [mcadGMeanError, thick] table [
     col sep=comma,
     x expr=\coordindex,
     y=GMean,
 ]{\coffeelaketable};
\end{axis}
\begin{axis}[
        ylabel near ticks,
        y tick label style={font=\scriptsize},
        y label style={font=\scriptsize},
        ylabel = {Diff. Throughput},
        xticklabels={,,},
        xtick pos=left,
        width=6.4cm,
        height=3.0cm,
        ymajorgrids=true,
        ytick={0.94, 0.96, 0.98, 1.0, 1.02},
        ymin=0.94,
        ymax=1.02,
    ]
    \addplot [mcadDTPPerf, thick, mark=diamond*, mark size=1.3pt, mark options=solid] table [
        col sep=comma,
        x expr=\coordindex,
        y=CoffeeLake,
    ]{\coffeelaketable};
    \addplot [mcadDTPMcad, thick, mark=o, mark size=1.3pt, mark options=solid] table [
        col sep=comma,
        x expr=\coordindex,
        y=MCAD,
    ]{\coffeelaketable};
\end{axis}
\end{tikzpicture}
\label{fig:clang-small-eval}
}
  \vspace*{-0.7em}
  \caption{Differential throughput prediction of small changes in \texttt{ffmpeg} and \texttt{Clang} on Intel CoffeeLake}
  \vspace*{0.5em}
\end{figure}

For our evaluation, we repeat the experiments detailed in Section~\ref{sec:eval-ffmpeg} for \texttt{ffmpeg} on 7 minor releases for version 4.2: 4.2.1, 4.2.2, 4.2.3, 4.2.4, 4.2.5, 4.2.6, and 4.2.7. The \#~LoCC between minor releases range from 132 to around 2000 lines with a geomean of 735 lines, whereas \#~LoCC between major \texttt{ffmpeg} releases evaluated in Section~\ref{sec:eval-ffmpeg} range from 128k to 390k lines with a geomean of 227k.
Figure~\ref{fig:ffmpeg-small-eval} shows the differential throughput results along with their error percentage.

We repeat the experiment for \texttt{clang}, for which we evaluate 6 minor 14.0 releases: 14.0.1, 14.0.2, 14.0.3, 14.0.4, 14.0.5, and 14.0.6. The \#~LoCC between these minor releases range from 76 to around 4k lines with a geomean of 1171 lines. In contrast, the \#~LoCC between major \texttt{clang} releases evaluated in Section~\ref{sec:eval-clang} range from 2.6M to 6M lines with a geomean of 3.6M lines.
Figure~\ref{fig:clang-small-eval} shows the differential throughput results as well as the respective their error percentage.

Both experiments show that \coolname{}'s predictions follow the hardware cycle counts closely, with only marginal error rate of less than 1\% on average and 1.5\% in the worst case.

\begin{table*}
  \small
  \renewcommand{\arraystretch}{1.25}
  \centering
  \begin{tabular}{@{}l c c c c c @{}}
      \toprule
      & uiCA~\cite{uica}
      & OSACA~\cite{laukemann2018automated}
      & Ithemal~\cite{mendis2019ithemal}
      & LLVM~MCA~\cite{llvmmca}
      & \textbf{\coolname{}} \\

      \midrule 
      execution time   & Timeout after 48h. & Exit w/error after 24h.  & Exit w/error after 2m. & 219.98s & \textbf{52.69s} \\
      \texttt{ffmpeg} results & \xmark         & \xmark        & \xmark        & \cmark        & \cmark \\
      \texttt{clang} results  & \xmark         & \xmark        & \xmark        & \cmark        & \cmark \\
      memory usage     & ~113GB         & N/A           & N/A           & 29.39GB       & \textbf{2.16GB} \\
      \midrule
      scales to \# of instrs.     & 10$\sim$20     & $\sim$40      & 10$\sim$20    & $\sim$1000    & \textbf{>1000000} \\
      mean error   & \textbf{3\%}~\cite{uica}  & $\sim$30\%~\cite{uica} & $\sim$5\%~\cite{uica} & $\sim$20\%~\cite{uica} & \textbf{\totalError{}} \\
      benchmark type   & basic block    & loop kernel   & basic block   & loop kernel   & \textbf{whole program} \\
      supported ISAs   & x86\_64 only   & x86\_64 only  & x86 \& x86\_64  & \textbf{$\sim$20 ISAs} & \textbf{$\sim$20 ISAs} \\

      handles branches & \xmark         & \xmark        & \xmark        & \xmark        & \cmark \\
      \bottomrule
  \end{tabular}
  \caption[Caption with FN]{Comparison of different tools to predict cycle counts of software across various dimensions. While the error metrics differ, we present error numbers as reported by the most recent work~\cite{uica} for completeness.\footnotemark}
  \vspace*{-2em}
  \label{tab:eval-comparison}
\end{table*}
\subsection{Scalability and Comparison}\label{sec:eval-scalability}
Besides accurate predictions that generalize across control-flow transfers, another major goal of \coolname{} is scalability.
In particular, we aim for \coolname{} to scale up with the complexity of real-world target programs.
In this section, we compare against four state-of-the-art throughput prediction and analysis approaches: OSACA~\cite{laukemann2018automated}, Ithemal~\cite{mendis2019ithemal}, uiCA~\cite{uica}, and LLVM~MCA~\cite{llvmmca}.
We present the results in Table~\ref{tab:eval-comparison}.

First, we focus on the benchmarks these tools used in their repositories or publications.
We compare the type and size of benchmark, as well as their supported target instruction sets.
All prior art operates on the individual basic block level with the exception of LLVM~MCA which also contains designated support for loop kernels.
However, in both cases instruction sequences usually consist of only 10$\sim$20 instructions at most.
On the other hand, \coolname{} was designed to work on real-world program traces containing \emph{millions} of instructions.
\coolname{} also supports most of the hardware architectures that QEMU and LLVM support, which amounts to nearly 20 different Instruction-Set Architectures~(ISAs).
In contrast, most other tools are highly architecture specific and only support x86\_64.

We further evaluate these tools using the same reference input and compare their performance with respect to execution time and memory consumption.
For this purpose, we collected the execution trace of a \texttt{ffmpeg} invocation using version 4.2 as described in Section~\ref{sec:eval-ffmpeg} storing the results into a file.
The resulting instruction stream consists of roughly 27 million x86\_64 instructions.
To give state-of-the-art approaches the benefit of the doubt we perform this experiment on an 80-core Intel Xeon E7-4870 machine, clocked at 2.4GHz, equipped with 198GB of RAM and the same amount of swap space, setting a 48-hour time limit on the execution.

As shown, OSACA and Ithemal did not finish this task: OSACA bailed out with failures related to loading hardware models after parsing the input file; Ithemal promoted an out-of-memory error from its DynamoRIO~\cite{bruening2004dynamorio} runtime before bailing out.
Similarly, uiCA could not finish within the time limit after consuming significant amount of memory.
Last but not the least, despite being able to finish, LLVM~MCA consumed significantly more time and memory compared to \coolname{}, demonstrating the effectiveness of our changes over standard MCA in our implementation.

\footnotetext{uiCA uses the Mean Absolute Percentage Error~(MAPE) to compare the error of a prediction against a single execution on a physical device, whereas we use the mean error of the predicted \emph{difference} in cycles between two executions.}

\section{Discussion}\label{sec:discussion}
Current state-of-the-art throughput prediction approaches either explicitly model or learn microarchitectural implementation details, resource usage, instruction scheduling, and latency numbers, using data.
This data is sometimes provided by processor vendors directly, although, most of the time it is collected using emperical methods, like measuring instruction latencies using many different software configurations, and large numbers of repetitions to reduce inherent error signals.

\coolname{} builds on top of this prior work that provides insights into modern processor pipelines through detailed measurements and experiments.
Since a lot of this research has been contributed in part by vendors directly and in other parts incorporated by the community into the LLVM compiler infrastructure, \coolname{} currently uses LLVM~MCA as the core analysis engine.
However, the core analysis component in our design can support other throughput analysis engines in principle, which would allow us to predict timing effects of a number of optimizations in modern processors that are not currently modeled by LLVM~MCA, such as including instruction prefetching and branch prediction which usually happen in the processor frontend.
The main obstacle towards that as demonstrated by our experiments in Section~\ref{sec:eval-scalability}, however, remains overcoming scalability issues of the related approaches.

Looking ahead to future work we anticipate that research into analysis of multi-process and multi-thread executions should be feasible within \coolname{} in principle.
As introduced in Section~\ref{sec:design} we collect execution traces using QEMU and a custom plugin and certain recently-added QEMU plugin interfaces would allow us to distinguish traces originating from different virtual CPUs at runtime.
Nevertheless, how to incorporate modern processors' concurrency models into current throughput prediction approaches remains an open research question.
It would also be possible to substitute QEMU for other methods of execution trace collection entirely: for instance, leveraging binary rewriting tools would enable us to insert instrumentations that report the executed instructions natively.

Last but not least, we believe a more scalable, intuitive, and interactive timeline or waterfall view could provide developers with more insights by visualizing resource dependencies among instructions, pointing towards potential avenues for improving continuous development of timing-sensitive code.

\section{Related Work}\label{sec:related}
In this section, we categorize the most relevant prior approaches and also compare them against \coolname{}.
To the best of our knowledge, none of the existing systems supports analyzing timing effects of programs in a timely manner that supports a development-driven workflow that can guide implementation changes with respect to timing-sensitive system behavior.

A large body of prior research focused on static prediction of worst-case timing behavior~\cite{sehlberg2006static, li2007chronos, falk2009optimal, lisper2014sweet, hardy2017heptane}.
However, reasoning about timing properties of arbitrary programs reduces to the halting problem in the general case and as a result approaches for calculating worst case execution time make strong assumptions such as an upper bound on the number of loop iterations, recursion depth, effects of memory accesses, and external I/O operations.
In practice, this means that the user of traditional tools has to provide upper bound information for all loop constructs, recursion, avoid indirect memory accesses through pointers, and avoid the use of I/O operations in analyzed parts of the code.
Ensuring proper and correct usage then typically requires dedicated build toolchains and environment setups, as well as expert knowledge about the analysis framework.
Moreover, such tools typically over-approximate cycle counts up to several orders of magnitude over physical hardware execution in order to remain sound, with several tools providing timing estimates in units of \emph{wall-clock time} rather than cycles~\cite{mezzetti2011industrial, abella2015wcet}.

More recently, a number of approaches~\cite{uica, laukemann2018automated, llvmmca, mendis2019ithemal} proposed throughput modeling of machine code using parametric models for accurate, yet fast throughput prediction.
Such approaches predict timing aspects of a particular instruction sequence of the target program rather than reasoning about the entire set of possible executions at once like prior static approaches do.
While their underlying parametric models require knowledge of key microarchitectural aspects such as port usage, instruction latencies, and other internal details that may not be publicly available, recent advances in machine learning showed that architecture dependence can be tackled to some extent by \emph{learning} model parameters from data~\cite{mendis2019ithemal}.
However, without ruling out the use of learning-based solutions, our evaluation show that currently such approaches are severely limited with respect to scalability, providing throughput estimates only for a handful up to a few hundred instructions at most, also lacking support for prediction across control-flow transfers.
In contrast, \coolname{} handles complex binary programs containing literally millions of individual instructions with near-native execution speeds.

Dynamic approaches aim at providing detailed and concrete timing analyses of the running software using concrete inputs.
There are two main flavors of dynamic timing analysis tools: either using physical hardware tracing or using architectural simulators.
Approaches using physical tracing execute the program on the target architecture and measure cycle counts directly using the facilities provided by the device~\cite{rapitimewcet2017whitepaper, linuxperf}.
While in theory this yields the most precise results and should also be reasonably fast, in practice this is often not the case: the target architecture might be a production system that is not readily available to the developer running the test and in a collaborative environment each team would require their own physical device to test their changes against.
Additionally, setting up and using facilities for accurate cycle-count measurements can be a time-intensive task in and of itself, requiring complicated setup, and potentially support by the target program's build toolchain as well as the operating system of the production system.

Cycle-accurate architectural simulators~\cite{burger1997simplescalar, yourst2007ptlsim, binkert2011gem5} on the other hand promise to provide a similar level of accuracy as physical tracing without requiring an actual physical device to capture program execution.
Unfortunately, simulation-based approaches also come with major drawbacks: first, performance is typically at least three orders of magnitude slower than native execution (or even slower) as they faithfully simulate microarchitectural details of modern processor pipelines completely in software.
Second, they are usually aimed towards explorative hardware design and implementation studies of novel architectures rather than simulating throughput of software for existing platforms.
As a result, these frameworks are not easily accessible and can be difficult to integrate with existing software development tools and continuous integration workflows due to the high resource requirements and time-intensive nature of the simulation-based approach.


\section{Conclusion}
To summarize, our results show that \coolname{} improves on the state of the art by scaling up to complex real-world software. It is well suited to providing cycle count estimates with rapid developer-centric turn-around times while targeting a range of different hardware architectures.
\coolname{} can drive software development by quickly iterating on small changes and assessing their timing impact on real-world programs such as \texttt{ffmpeg} and \texttt{clang}, with a mean error in differential throughput estimates of $<$~\totalError{} compared to hardware-based measurements.



\bibliographystyle{ACM-Reference-Format}
\bibliography{refs}

\end{document}